\documentclass[12pt]{book}
\usepackage{fullpage}

\usepackage{threeparttable}
\usepackage{lscape}

\usepackage{graphicx}
\usepackage{amsmath}
\usepackage{longtable}
\usepackage[none]{hyphenat}
 \usepackage{setspace}
 \usepackage{url}

\usepackage{lineno}
\newcommand{\Dmel}{\emph{D. melanogaster}}

\newcommand{\Dana}{\emph{D. ananassae}}
\newcommand{\Dyak}{\emph{D. yakuba}}

\newcommand{\Dsim}{\emph{D. simulans}}

\newcommand{\Dros}{\emph{Drosophila}}

\newcommand{\fiveP}{5$'$}
\newcommand{\thrP}{3$'$}

\newcommand{\Rebekah}{Rebek\mbox{}ah }

\makeatletter
\renewcommand{\@biblabel}[1]{\quad#1.}
\makeatother
\usepackage{natbib}
\bibpunct[; ]{(}{)}{,}{a}{}{;}

        \usepackage{hyperref} 

        \hypersetup{ colorlinks=true,
                     linkcolor=blue, 
                     citecolor=blue, 
                     filecolor=blue, 
                     urlcolor=blue 
                   }
\usepackage{textcomp}

\begin{document}

\author{\Rebekah L. Rogers$^1$, Ling Shao$^1$, Jaleal S. Sanjak $^1$, Peter Andolfatto$^2$, \\ and Kevin R. Thornton$^1$}

\title{ Revised annotations, sex-biased expression, and lineage-specific genes in the \emph{Drosophila melanogaster} group}
\date{}

{\hfill \textbf{}
\let\newpage\relax\maketitle}

\begin{center} \Large Research Article \end{center}
\vspace{0.25in}

\noindent 1) Ecology and Evolutionary Biology, University of California, Irvine \\
\noindent 2)  Ecology and Evolutionary Biology and the Lewis Sigler Institute for Integrative Genomics, Princeton University \\

\vspace{0.25in}

\noindent \textbf{Running head: } Gene annotations and sex-biased expression in \Dros

\vspace{0.25in}


\vspace{0.25in}

\noindent \textbf{Corresponding author:} Rebekah L. Rogers,  Dept. of Ecology and Evolutionary Biology, 5323 McGaugh Hall, University of California, Irvine, CA 92697 \\
\\
\noindent \textbf{Phone:}  949-824-0614

\noindent \textbf{Fax:}  949-824-2181

\noindent \textbf{Email:} rogersrl@uci.edu

\chapter*{}
\section*{Abstract}
Here, we provide revised gene models for \Dana, \Dyak, and \Dsim, which include UTRs and empirically verified intron-exon boundaries, as well as ortholog groups identified using a fuzzy reciprocal-best-hit blast comparison.   Using these revised annotations, we perform differential expression testing using the cufflinks suite to provide a broad overview of differential expression between reproductive tissues and the carcass.  We identify thousands of genes that are differentially expressed across tissues in \Dyak {} and \Dsim, with roughly 60\% agreement in expression patterns of orthologs in \Dyak {} and \Dsim.  We identify several cases of putative polycistronic transcripts, pointing to a combination of transcriptional read-through in the genome as well as putative gene fusion and fission events across taxa.  We furthermore identify hundreds of lineage specific genes in each species with no blast hits among transcripts of any other \Dros {} species, which are candidates for neofunctionalized proteins and a potential source of genetic novelty.

\section*{Introduction}

Accurate models of gene structure including UTRs, intron-exon boundaries, as well as coding sequences are essential for proper interpretation of molecular genetics \citep{Fire1998, CRSPR}, demographic inference \citep{Halligan2006, Parsch2010, Clemente2012}, tests of selection \citep{McDonald1991}, and comparative genomics \citep{modENCODE}.  The \Dros {} offer an excellent model for comparative genomics, with high-quality sequenced genomes for 12 species\citep{TwelveGenomes} as well as draft genomes for an additional 8 species  \citep{modENCODE} spanning a total of 63 MY \citep{Tamura2004}.  Previous gene models provided for the 12 \Dros {} genomes focused on gene prediction with the aid of homology to establish putative annotations of coding sequences across taxa with 15,000-16,000 genes for most species \citep{TwelveGenomes}.  These gene models produce reliable annotations for conserved genes as well as genes that are present in multiple species but will lack lineage specific genes where sources of genetic novelty add to the genome, will misidentify rapidly evolving sequences, and will neglect isoforms that may offer alternative molecular functions.  
 
Recent work has validated gene models in \Dmel {} through cross-species comparisons \citep{modENCODE}.  While aligned CDS sequences often display patterns of expression consistent with conservation, gene structure varies across taxa at UTRs, introns, and noncoding DNA \citep{modENCODE}.  Furthermore, any gene families and functional classes subject to rapid evolution in gene structure which is unlikely to be reflected in homology based annotations \citep{Wasbrough2010} and \emph{de novo} genes will also be absent in spite of their role in developing functional diversity \citep{Zhao2014}.   

Beyond the evolution of gene structure, expression patterns of genes across tissues is expected to influence their evolutionary constraints and rates of sequence evolution \cite{VanDyken2010} as well as their functional roles within the organism.  Sex-biased genes are often rapidly evolving \cite{Ranz2003,Ellegren2007} especially among male reproductive proteins \cite{Zhang2007, Meisel2011, Haerty2007}.  Moreover, sex specific expression patterns can influence the extent to which genes contribute to sexual antagonism, influencing their role in the evolution of sexual dimorphism \cite{Mank2008}.  Correctly identifying sex-specific and tissue specific expression is therefore essential for studying the molecular and evolutionary impacts of genes within species.  Moreover, changes in tissue specific or sex-specific expression may point to rapid evolution of gene functions and candidate loci to search for signals of evolutionary change and differential selective effects between the sexes.

Here, we describe RNA-seq based gene annotations for \Dros {} outgroup species \Dsim, \Dyak, and \Dana {} based on whole transcriptome sequencing of male and female adult tissues.   These revised gene models capture hundreds of lineage specific genes on major chromosomal arms in \Dyak {} and \Dsim.  We also describe \fiveP {} and \thrP {} UTRs, and intron-exon structure for genes throughout the \Dmel {} group.   Finally, we describe sex-biased expression across species, identifying thousands of genes that are differentially expressed across tissues.   These revised gene models as well as results of sex-biased and tissue-biased differential expression testing should serve as a resource for the \Dros {} evolutionary and molecular genetics community interested in evolution, conservation, and gene expression.

\section*{Methods}
\subsection*{Sample preparation}
Fly stocks were incubated under controlled conditions at 25\textcelsius {} and $\sim$40\% humidity.  Virgin flies were collected within 2 hours of eclosion, then aged 2-5 days post eclosion before dissection.  We dissected samples in isotonic Ringers solution, using female ovaries and headless gonadectomized carcass from two adult flies as well as testes plus glands and male headless gonadectomized carcass for four adult flies for each sample RNA prep.  We collected three biological replicates of the \Dyak {} reference, three biological replicates of the \Dsim {} $w^{501}$ reference, and one replicate of the \Dana {} reference and one replicate of the \Dmel {} reference (stock numbers in Table \ref{FlyStocks}).  Samples were flash frozen in liquid nitrogen immediately after dissection, and and stored in 0.2ml Trizol at -80\textcelsius.  All samples were homogenized in 0.5ml Trizol Reagent (Invitrogen) with plastic pestle in 1.5ml tube, mixed with 0.1ml chloroform, and centrifuged 12,000g 15min at 4oC, as Trizol RNA extraction protocol. The RNAs in the supernatant about 0.4ml were then collected and purified with Direct-Zol RNA MiniPrep Kit (Zymo), following the standard protocol. The total RNAs were eluted in 65$\mu$L  RNase-Free H$_2$O.  About 1$\mu$g purified RNAs were treated with 2$\mu$L Turbo DNase (Invitrogen) in 65$\mu$L  reaction, incubated 15min at room temperature with gentle shaking. These RNAs were further purified with RNA Clean and Concentrator-5 (Zymo). One extra wash with fresh 80\% ethanol after the final wash step was added into the original protocol. The treated RNAs were eluted with 15$\mu$L RNAse-Free  H$_2$O, and stored at -80\textcelsius. 

The amplified cDNAs were prepared from 100ng DNase treated RNA with Ovation RNA-Seq System V2 (Nugen) and modified protocol. The preparations followed the protocol to the step of SPIA Amplification (Single Primer Isothermal Amplification). The amplified cDNAs were first purified with Purelink PCR Purification Kit (Invitrogen, HC Binding Buffer) and eluted in 100$\mu$L  EB (Invitrogen). These cDNAs were purified again to 25$mu$L EB with DNA Clean and Concentrator -5 Kit (Zymo) for Nextera library preparation.  About 43ng cDNAs were used to construct libraries with Nextera DNA Sample Preparation Kit (Illumina) and modified protocol. After Tagmentation, Purelink PCR Purification Kit with HC Binding Buffer was used for purification and eluted with 30$\mu$L  EB or H2O. The products (libraries) of final PCR amplification were purified with DNA Clean and Concentractor-5 and eluted in 20$\mu$L EB. The average library lengths about 500bp were estimated from profiles on a Bioanalyzer 2100 (Agilent) with DNA HS Assay. All libraries were normalized to 2-10nM based on real-time PCR method with Kapa Library Quant Kits (Kapa Biosystems).  The qualities and quantities of these RNAs, cDNAs and final libraries were measured from Bioanalyzer with RNA HS or DNA HS Assays and Qubit (Invitrogen) with RNA HS or DNA HS Reagents, respectively.  Samples were barcoded and sequenced using 76 bp reads in 4-plex on an Illumina HiSeq 2500 using standard Illumina barcodes, resulting in high coverage (Figure \ref{DsimFemCov}-\ref{DmelCov}).   Coverage in RNA-seq data is dependent on expression level, but the majority of sites have coverage less than 50 reads.  One replicate of female tissues was sequenced using single end reads.  All other libraries were sequenced using paired end reads and dual indexing.  Samples were sequenced as they became available, across the equivalent of one eight lane flow cell for the 32 samples in total. 
 
\subsection*{Trinity and Augustus gene annotation}
RNA sequencing data for ovary, female carcass, testes, and male carcass of \Dyak, \Dsim, and \Dana {} were concatenated into a single fastq file, and digitally normalized to remove excess redundant reads for highly expressed transcripts.  This step results in tractable runtimes with no loss of information in transcript annotations. We ran Trinity \url{http://trinityrnaseq.sourceforge.net/}\citep{Haas2013,Grabherr2011}. 
For a single sample:

\noindent1.  FASTQ files for left and right reads were concatenated.
\noindent2.  The concatanated files were subject to "digital normalization" using the following command in the Trinity package:
   
\noindent\verb!normalize_by_kmer_coverage.pl --seqType fa --JM 100G --max_cov 30 --left left.fa! \\ \noindent\verb!--right right.fa --pairs_together --PARALLEL_STATS --JELLY_CPU $CORES!, where \$CORES is how many CPU we had available

\noindent3.  The resulting normalized left and right fastq files where then used in the following command: \verb!    Trinity.pl --seqType fa --bflyHeapSpaceInit 1G --bflyHeapSpaceMax 8G!\\ \noindent\verb!--JM 7G --left $LEFTFILE --right $RIGHTFILE --output trinity_output! \\ \noindent\verb!--min_contig_length 300 --CPU $CORES --inchworm_cpu $CORES --bflyCPU $CORES!,  where the variables are the normalized fastq files and the number of cores available.

\noindent4.  Further detail is at \url{https://github.com/ThorntonLab/annotation_methods}

  The resulting annotations were used as input in the Augustus v2.5.5 gene prediction software \url{http://bioinf.uni-greifswald.de/augustus/} with command line options \verb!--species=fly --hintsfile=$INFILE_BASE.hints.E.gff !\\ \noindent\verb!--extrinsicCfgFile=augustus.2.5.5/config/extrinsic/extrinsic.ME.cfg !\\ \noindent\verb!$INFILE --gff3=on --uniqueGeneId=true > $INFILE_BASE.gff3!.

\subsection*{Alignment and annotation}

We matched Augustus gene models to previous annotations from FlyBase r1.3 for \Dana, and \Dyak, or to the \Dsim {} $w^{501}$ reference annotations.  CDS sequences were required to physically overlap with the location of a current FlyBase or $w^{501}$ gene model and were required to have matches to 85\% or more CDS sequence with 90\% or greater amino acid similarity in an all-by-all BLASTp of translated sequences at a cutoff of $E\leq 10^{-10}$ with low-complexity filters turned off (-F F).   We mapped RNA-seq data to known gene models annotated in FlyBase, \Dyak {} r1.3, \Dana {} r1.3, \Dmel {} r.5.45, and the \Dsim {} $w^{501}$ gene models annotated by Hu et al. \citep{SimRef}.   GFF Files were reformatted using gffread from Cufflinks suite.  Sequences were mapped to the genome using Tophat v.2.0.6 \citep{Trapnell2009,Kim2013} and Bowtie2 v.2.0.2 \citep{Langmead2009}, using reference annotations as a guide, with no attempt to identify novel transcripts using reads which fell outside reference annotations (-G) and all other parameters set to default.    We used Cufflinks 2.0.2 \citep{Trapnell2012} to calculate expression levels across genes and transcripts, normalizing expression by the upper quantile (-N) and ignoring reads which fall outside known gene models (-G) with all other options set to default.  Orphaned gene models from FlyBase or from Hu et al. (2012) which had FPKM $\geq 2$ but which had no assembled gene model match from Augustus were included in the final annotations used for differential expression testing.  Some annotations contain polycistronic transcripts encompassing multiple independent open reading frames.  A portion of these polycistronic transcripts may reflect only low-level polycistronic transcription, rather than polycistronic transcripts serving as the dominant isoforms but the rate of polycistronic transcription cannot be readily determined with available data.  For genes with polycistronic transcripts but no 1:1 transcript match with FlyBase gene models, we included annotations for both the polycistronic Augustus gene models and the gene models from FlyBase supporting independent transcripts.  

\subsection*{Sex-bias and Tissue Bias}
The union of gene models from Augustus and orphaned gene models from FlyBase were combined into a single GFF containing transcript and CDS annotations for each species.  We then re-mapped RNA-seq reads to gene models in the reannotated GFF for \Dyak, \Dana, and \Dsim, as well as unmodified gene models for \Dmel {} r.5.45 with Tophat and performed differential expression testing at an FDR $\leq$ 0.1 normalizing expression by the upper quantile (-N) and ignoring reads which fall outside known gene models (-G) with all other options set to default using the Cufflinks suite according to the same criteria described above.  We compared female carcass to female ovaries, male carcass to male testes, female carcass to male carcass, and female ovaries to male testes for each species, grouping replicates for reference genomes.      

\subsection*{Orthologs and lineage specific genes}
Orthologs were identified using fuzzy reciprocal best hit BLASTp comparisons of all translations across reference genomes for gene model predictions of \Dana, \Dyak, \Dsim, as well as \Dmel.  Orthologs are similar to those previously used to annotate gene families in \Dros {} \citep{TwelveGenomes, Hahn2007}.  Putative orthologs must be putative reciprocal hits of the same rank order, where genes with an E-value within a single log-unit of one another are assigned the same rank, using the best E-Value for a gene with a cutoff of $E\leq10^{-10}$.  Lineage specific genes were defined as genes with no hit in an all-by-all BLASTp of translations against translations of the other outgroups (e.g. \Dyak, \Dana, and \Dmel {} for \Dsim) at a cutoff of $E\leq10^{-10}$ with low-complexity filters turned off (-F F).

\subsection*{Gene ontology}
We used DAVID gene ontology analysis software \url{http://david.abcc.ncifcrf.gov/} to determine whether any functional categories were overrepresented among genes with sex specific or tissue specific expression.  Functional data for \Dana, \Dyak {} and \Dsim {} are not readily available in many cases, and thus we identified functional classes in the \Dmel {} orthologs as classified in Flybase. Gene ontology clustering threshold was set to Low.    Genes with tissue specific expression were based on genes with differential expression from cufflinks for comparisons of female carcass vs. female ovaries and male carcass vs. male testes at a genomewide FDR $\leq 0.10$, according to cufflinks default settings.

\section*{Results}
\subsection*{Annotations}
\begin{figure}
\includegraphics[scale=0.4]{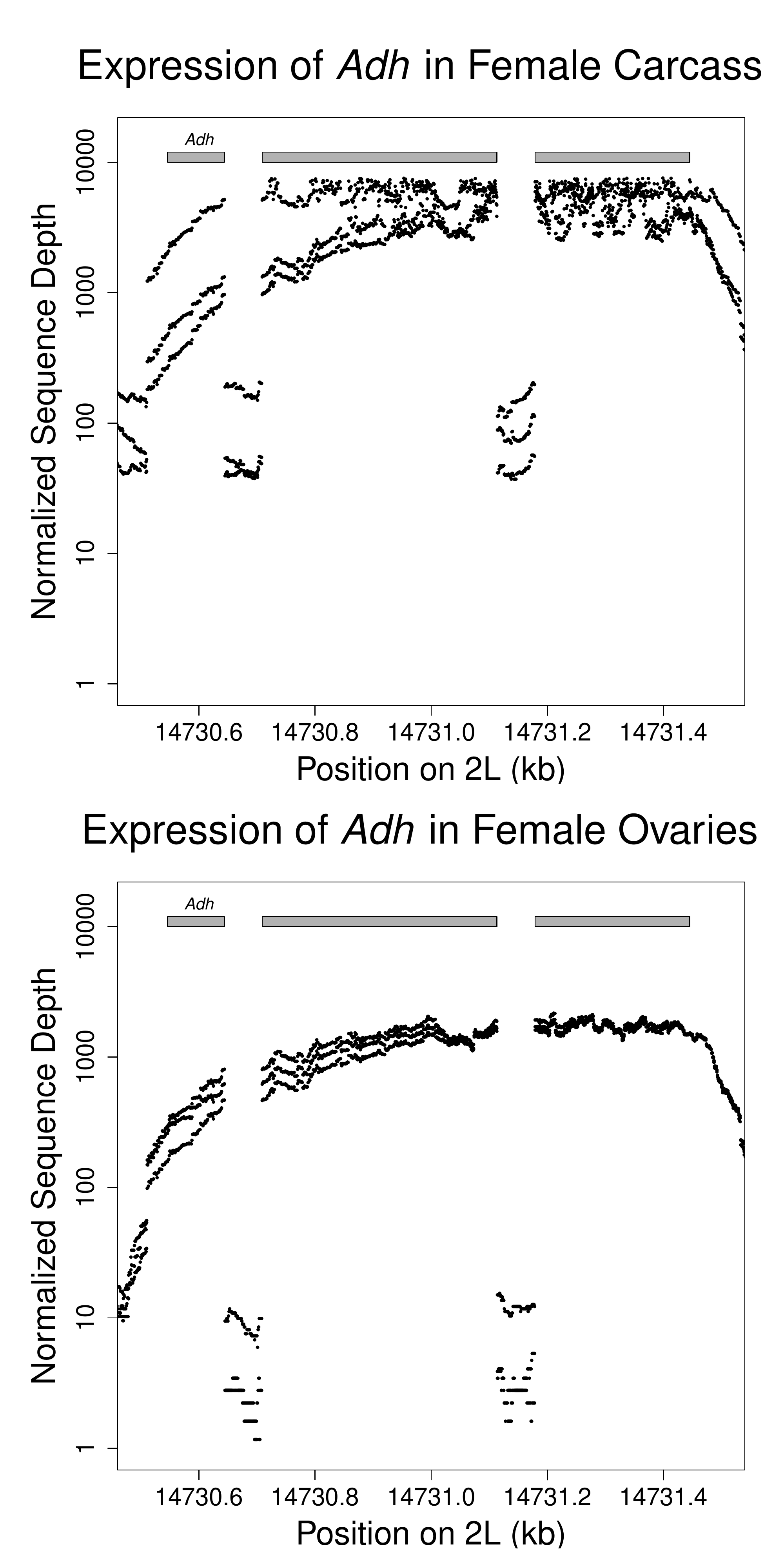}
\caption{\label{AdhFig}Quantile normalized coverage for reference strains at the \emph{Adh} locus in \Dyak. Coverage shows clear distinctions between introns and exons, and coverage that spans both \fiveP {} and \thrP {} UTRs in ovaries and carcass.  Low coverage of intron sequence points to partial sequencing of low levels of unprocessed transcripts.  }
\end{figure}

\begin{figure}
\includegraphics{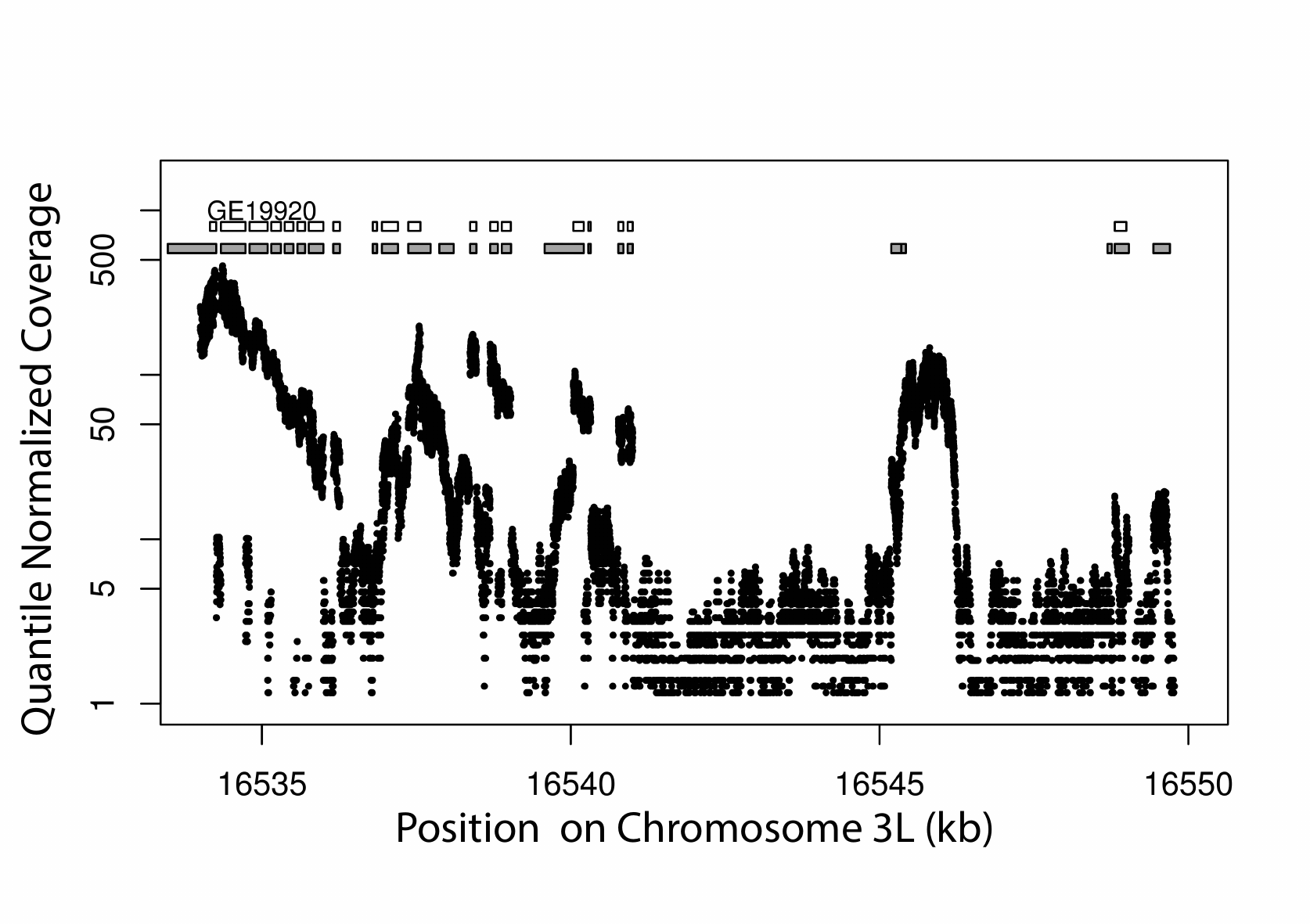}
\caption{\label{AnnotationFig} Quantile normalized RNA-seq data for three replicates of the \Dyak {} reference with an example of flybase gene model (white) and revised gene model (grey).  }

\end{figure}

For moderately to highly expressed genes we recover gene structure with intron-exon boundaries and UTR sequences for full length transcripts including novel exons which were previously unannotated based on comparative genomics (Figure \ref{AdhFig}-\ref{AnnotationFig}).   Many genes are part of polycistronic transcripts including one from \Dmel.  We identify 2529 putative polycistronic transcripts in \Dyak, 2379 in \Dana, and 561 in \Dsim.  For such genes we offer gene models from FlyBase as well as fused models from Augustus.  The extent to which such transcription of multiple genes is functional as opposed to a stochastic byproduct of transcriptional errors is unclear.  We also include FlyBase gene models expressed in the reference genomes with no 1:1 match in gene models from Augustus.   The addition of FlyBase gene models results in an additional 4265 annotations in \Dyak, 5367 in \Dana, and 7419 in \Dsim.  We identify a total of 22,989 transcripts for 16,278 genes in \Dsim, 20,315 transcripts for 17,579 genes in \Dyak, and 22,420 transcripts for 20,580 genes in \Dana {} (Table \ref{GeneCounts}). Compared to \Dyak, for \Dana {} an additional 1,173 FlyBase gene models failed to match sufficiently well with RNA-seq supported gene models and these were added to the annotations, explaining some of the excess in the number of genes and also highlighting the difficulties of annotation through comparative genomics across large phylogenetic distances.  In \Dana, 72\% of gene models have RNA-seq data supporting 60\% or more gene features (exons, UTRs) compared to 79.4\% in \Dyak {} and 80.1\% in \Dsim {} (Table \ref{PercentSupport}).

As defined by a fuzzy reciprocal best-hit blast \citep{TwelveGenomes, Hahn2007} we identify 12,127 genes in \Dmel {} with first-order orthologs in \Dsim {}, 11,425 with first-order orthologs in \Dyak, and 11,348 with first-order orthologs in \Dana {} (Table \ref{AllOrthoCounts}).  The increase in the number of genes with orthologs in \Dsim {} is the product of improved annotations as well as the improved assembly of the $w^{501}$ \Dsim {} reference\citep{SimRef}.   We observe a 1:1 concordance for 48\% of FlyBase gene models in \Dyak {} and 46\% of FlyBase gene models for \Dana.  These annotations typically include UTR sequences in addition to empirically supported intron-exon and coding sequence  boundaries, an improvement over previous gene models from release r1.3 which lack UTRs \citep{TwelveGenomes}.   We further identify thousands of lineage specific genes in each species of \Dros {} with no matching gene model in other outgroup species.  While lineage specific genes identified on minor chromosomes could result from assembly issues, we identify hundreds on major chromosomal arms (Table \ref{NewGenes}) suggesting that many of these are in fact cases of lineage-specific gene formation.

\subsection*{Sex-biased and tissue-biased expression}
\begin{figure}
\includegraphics{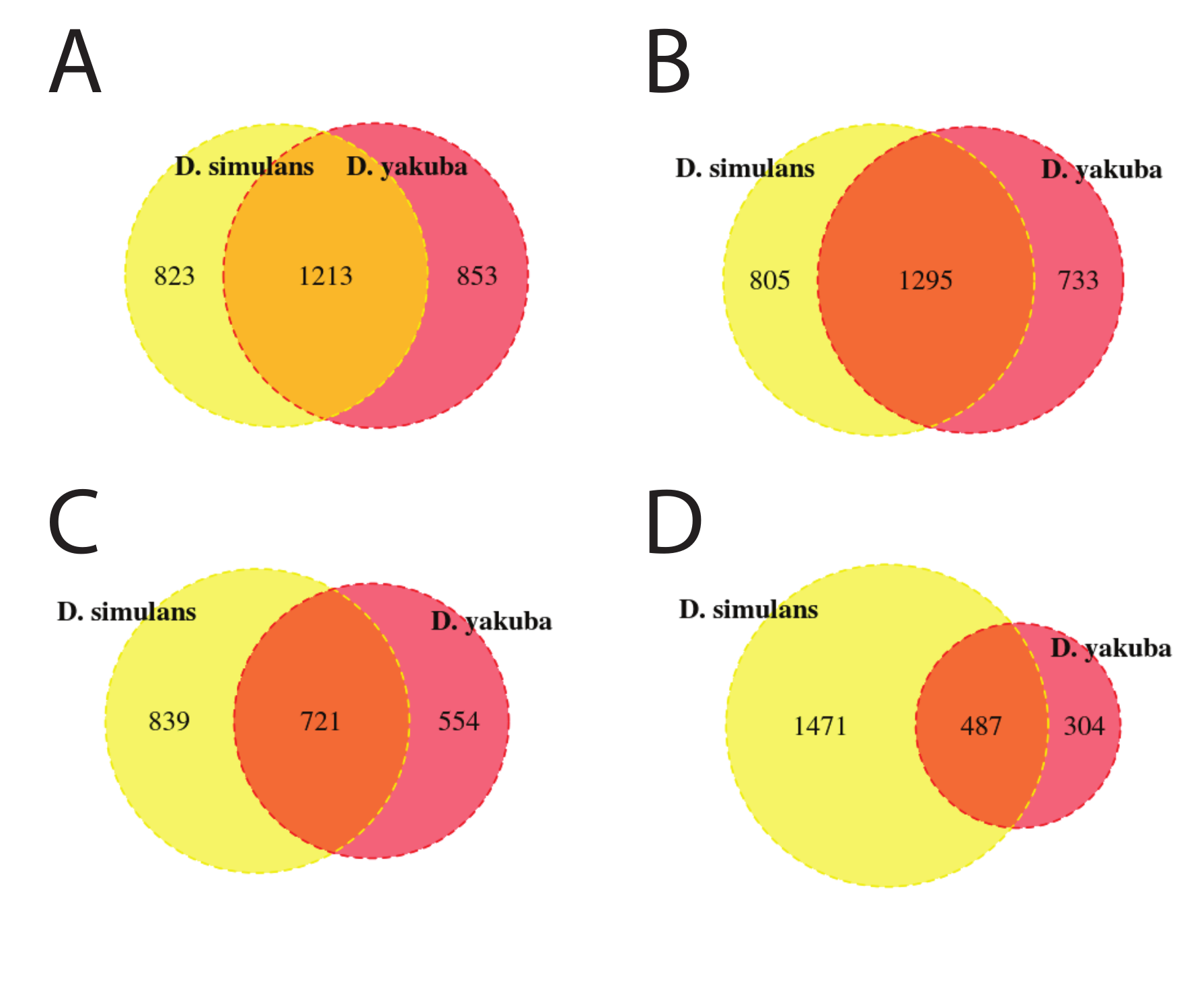}
\caption{\label{VennDiagrams} Genes with tissue biased expression in both \Dyak {} and \Dsim {} in A) female ovary B) female carcass C) male testes D)  male carcass.  Numbers shown include only genes with a reciprocal best hit ortholog in the sister species.   }

\end{figure}

We observe thousands of genes with sex-biased or tissue-biased expression in \Dyak {} and \Dsim, but hundreds of genes with sex-biased or tissue-biased expression in \Dmel {} and \Dana, a direct product of increasing power to detect differences in RNA levels with biological replicates.   Gene ontology categories overrepresented between ovary and female carcass reflect differences in genes involved in reproduction, chromosome segregation and DNA synthesis or repair, while genes differentially expressed between testes and carcass reflect sperm development, cell division, and energy production (see Supplementary Information).  We used reciprocal best hit orthologs to identify genes with similar regulation across species, focusing on \Dyak {} and \Dsim {} where biological replicates increase power for differential expression testing (Table \ref{DiffExpGeneTab}).  A total of 10,369 genes in \Dyak {} have reciprocal best hit orthologs in \Dsim, and were retained to compare differential expression across the two species.   We have collected replicates for both \Dyak {} and \Dsim, contributing to the greater power in differential expression testing.   Roughly 60\% of genes with tissue biased expression in \Dyak {} that have a reciprocal best hit ortholog in \Dsim {} exhibit the same tissue specific bias in \Dsim, with marginally greater agreement in genes biased toward the carcass than the reproductive tissues in both males and females (Figure \ref{VennDiagrams}).  We additionally observe evidence of differential expression in at least one of the four comparisons of male and female germline or somatic tissues for 118 lineage specific genes in \Dana, 334  in \Dyak, and  222 in \Dsim {} (Table \ref{NewGenes}), suggesting that they are not solely artifacts of gene annotation software.

\section*{Discussion}
\subsection*{Gene models and ortholog calls}

Correct interpretations of gene expression changes and gene family evolution depend on accurate gene models.  Among the \Dros, \Dmel {} has received the most attention with empirically verified gene annotations through high throughput EST and RNA-seq data as well as detailed manual or molecular curation of single genes. Until present, gene models for outgroup species offer only CDS sequences, with no information concerning \fiveP {} or \thrP {} UTRs or alternative isoforms even for well-studied genes such as \emph{Adh}.  Establishing more complete gene models based on RNA-seq data will allow us to correctly identify coding and non-coding sequences during functional assays, correctly identify putatively neutral vs non-neutral mutations, and correctly define new mutations including the origins of new genes, expansion of gene families, and gross modification of coding sequences through rearrangement, duplication, and deletion.  Moreover, identifying putative isoforms provides a more complete portrait of gene structure and function across species.   Here, we provide updated gene annotations based on high coverage RNA-seq data for the reference genomes of 3 species of \Dros: \Dana, \Dyak, and \Dsim, which should serve as an excellent springboard and initial resource to the \Dros {} molecular and evolutionary genetic community.  

Additionally, lineage specific genes are unlikely to be identified in previous annotation efforts that focused largely on conserved amino acid sequences.  With RNA-seq annotations we can identify lineage specific genes, which are expected to be important in the evolution of genomes and emergence of genetic novelty.  The increased power to identify genes independently from conservation is a major step toward studying the evolution of genome content and rapidly evolving gene sequences.  The ability to identify transcribed genes independently from conservation will facilitate evolutionary and functional analyses of the most rapidly changing segments of the genome.    There may be additional gene models and isoforms in other tissue types or timepoints which deserves to be explored.  Particularly, these annotations are unlikely to reflect the full diversity of sequences and isoforms expressed during embryonic development post-fertilization, pupal development, or in larvae.  These alternative timepoints deserve future exploration and the gene models offered here will serve as a lower bound on the full diversity of sequences that are transcribed within each species.  
 
\subsection*{Polycistronic genes}
We observe thousands of putatively polycistronic annotations in \Dana {} and \Dyak, as well as hundreds in \Dsim {} $w^{501}$ where fewer gene model annotations were previously aligned and power to identify polycistronic genes is limited.  In \Dmel, hundreds of genes are known to show signs of polycistronic transcription \citep{Lin2007} offering a means of co-regulation across genes with similar functions \citep{Slone2007,Blumenthal1998}.  With very high sequencing coverage we are able to recover a greater number of polycistronic transcripts, though some of these may be false positives resulting from annotation algorithms.  Some genes may differ in the frequency with which they are transcribed as polycistronic vs. independent transcripts, but these results imply that at least low levels of polycistronic transcription may be common for many genes in the genome and future validation  may explore the extent of their functionality.    \emph{Adh} and \emph{Adhr} show evidence of differing polycistronic status across \Dros {} species \citep{Betran2000} and plant genomes are thought to split and fuse genes at rates of roughly $10^{-11}-10^{-10}$ per gene per year \citep{Nakamura2007}.   Switching the rates at which genes are co-transcribed has the potential to alter regulatory patterns across species \citep{Slone2007,Blumenthal1998} and thereby produce novel phenotypes.  These differences in polycistronic transcription therefore represent potential sources of genetic change that may be important in evolutionary change.

\subsection*{Tissue specific expression}
We identify thousands of genes that are differentially expressed across tissues in \Dyak {} and \Dsim, where biological replicate samples for each tissue are available.  A large fraction of the genome appears to display tissue biased expression between germiline and carcass, with many fewer genes showing differential expression between male and female gonadectomized carcass, consistent with early microarry-based assays in \Dmel {} \citep{Parisi2003,Parisi2004}.  We have sequenced samples to extremely high coverage, generating transcript annotations with \fiveP {} and \thrP {} UTRs with empirically supported intron-exon structures, improving accuracy in differential expression testing.  Biological replicates for \Dyak {} and \Dsim {} result in much greater power to identify differentially expressed genes in comparison with \Dana {} and \Dmel.   However, even in \Dana {} and \Dmel {} where only one replicate was available per tissue, we are still able to identify hundreds of genes that are differentially expressed across tissues with high coverage.  
 
 A comparison of orthologs between \Dyak {} and \Dsim {} where biological replicates should result in sufficient power to detect differential expression reliably across both species reveals that roughly 60\% of genes with reciprocal best hit orthologs exhibit similar tissue biased expression between the two species.  The remaining 30\% represent either genes that are differentially regulated between tissues but with effect sizes beyond the limits of detection, or genes that have evolved independent expression patterns between the two species.  We also observe tissue biased expression in hundreds of lineage specific genes, which may represent candidates for neofunctionalization and new gene origination.   These genes that display changes in sex-biased or tissue-biased expression across taxa, as well as lineage specific genes that exhibit tissue-biased and sex-biased expression are important candidate loci for evolutionary change in genome content and function that will be useful in exploring the functional and selective impacts of genomic changes.

\section*{Data access}
Gene annotations, ortholog calls, gene ontology calls, and Cuffdiff differential expression testing output files for all samples are available at \url{http://github.com/ThorntonLab/GFF}.  RNA-seq based annotations as well as first order orthologs in comparison to \Dmel {} can be viewed in the UCSC browser on the Thornton Lab public track hub at  \url{http://genome.ucsc.edu}. Sequencing fastq files were deposited in SRA with accession numbers PRJNA196536, PRJNA193071, PRJNA257286, and PRJNA257287.  

\section*{Author Contributions}
RLR, PA, and KRT designed experiments.  LS and RLR performed experiments and collected data.  RLR, JS, and KRT performed analyses. RLR and KRT wrote the manuscript with methodological input from LS.  

\section*{Acknowledgements}
We would like to thank Rahul Warrior for offering incubator space for fly cultures and Rachel Martin for an emergency supply of liquid nitrogen.  RLR is supported by NIH Ruth Kirschstein National Research Service Award F32-GM099377.  Research funds were provided by NIH grant R01-GM085183 to KRT and R01-GM083228 to PA.  All sequencing was performed at the UC Irvine High Throughput Genomics facility, which is supported by the National Cancer Institute of the National Institutes of Health under Award Number P30CA062203.  The content is solely the responsibility of the authors and does not necessarily represent the official views of the National Institutes of Health.  The funders had no role in study design, data collection and analysis, decision to publish, or preparation of the manuscript.

\bibliographystyle{Genetics}
\bibliography{NewAnnotations}
\clearpage
\begin{table}
\begin{center}
\caption{\label{FlyStocks}Fly stocks used for RNA-seq}
\begin{tabular}{ll}
Species & Strain \\
\hline
\Dmel{} & 14021-0231.36  \\
\Dsim {} &  $w^{501}$  \\
\Dyak {} & 14021-0261.01  \\
\Dana {} & 14024-0371.13  \\
\hline
\end{tabular}
\end{center}
\end{table}

\clearpage

\begin{table}

\caption{\label{GeneCounts}Number of transcripts and genes identified}
\begin{center}
\begin{tabular}{llrrrr}
Species & & Transcripts & & Genes & \\
\hline
 &  Previous Release & Revised  & FlyBase  & Revised   & FlyBase   \\
\hline
\Dsim  & r1.3  &18,781 & 15,415 & 16,278 & 15,413 \\
\Dyak  &r1.3 &  20,239  &16,082  & 17,579 &16,077 \\
\Dana  & r1.3 &  22,418 & 15,070& 20,580  & 15,069 \\
\hline
\end{tabular}
\end{center}
\end{table}
\clearpage

\begin{table}

\caption{\label{PercentSupport}Percent of revised gene models with $\geq$ 60\% of features supported by RNA-seq data}
\begin{center}
\begin{tabular}{lr}
Species &  Percent Suported \\
\hline
\Dana & 72.3\%  \\
\Dyak & 79.4\%  \\  
\Dsim & 80.05\% \\
\hline
\end{tabular}
\end{center}
\end{table}

\clearpage
\begin{table}

\caption{\label{AllOrthoCounts}Genes with a first order ortholog identified}
\begin{center}
\begin{tabular}{llrr}
Genes in & With an ortholog in & Revised & FlyBase  \\
\hline
\Dmel & \Dsim & 12,199 & 10,705\\
\Dmel & \Dyak & 11,472 & 11,556\\
\Dmel & \Dana & 11,451& 10,938 \\
\hline
\Dsim & \Dmel &  13,295 & - \\
\Dsim & \Dyak &  12,299 & - \\
\Dsim & \Dana &  11,994 & - \\
\hline
\Dyak & \Dmel &  12,868  & - \\
\Dyak & \Dsim & 12,831 & -  \\
\Dyak & \Dana & 12,337 & - \\
\hline
\Dana & \Dmel & 12,897 & - \\
\Dana & \Dsim & 12,612 & - \\
\Dana & \Dyak & 12,723 & - \\

\hline
\end{tabular}
\end{center}
\end{table}

\clearpage

\begin{table}

\caption{\label{NewGenes}Putative Lineage Specific Genes on Major Chromosomes}
\begin{center}
\begin{tabular}{lcrc}
Species &  Major chromosomes & Total & Diff Exp \\
\hline
\Dana & -  & 2977 &  118  \\
\Dyak & 230 & 1340 & 334  \\  
\Dsim & 369 & 1314 & 222\\
\hline
\end{tabular}
\end{center}
\end{table}

\clearpage
\begin{table}
\begin{center}
\caption{\label{DiffExpGeneTab} Differentially Expressed Genes By Tissue and Species  at FDR $\leq$ 0.1}

\begin{tabular}{lllrr}
Species & Tissue & Tissue & Significant  & Tested \\
\hline
\Dana {} & Female Ovary & Female Carcass &203 & 7537 \\
& Female Ovary & Male Testes & 200 & 8282  \\
&  Male Carcass & Male Testes & 1013 & 8349  \\
& Male Carcass & Female Carcass & 175 & 8417\\
\hline
\Dyak {} & Female Ovary & Female Carcass & 5420 & 8689 \\
& Female Ovary & Male Testes & 5868 & 10202\\
&  Male Carcass & Male Testes & 3065 & 10412  \\
& Male Carcass & Female Carcass &  724 & 9430 \\
\hline
\Dsim{} & Female Ovary & Female Carcass & 5053 & 8967 \\
& Female Ovary & Male Testes & 5741& 10222  \\
&  Male Carcass & Male Testes & 4628 & 10679 \\
& Male Carcass & Female Carcass & 611& 9566\\
\hline
\Dmel {} & Female Ovary & Female Carcass & 112 & 11326\\
& Female Ovary & Male Testes & 370 & 12890  \\
&  Male Carcass & Male Testes & 220 & 13268  \\
& Male Carcass & Female Carcass & 286 & 12502  \\
\hline
\end{tabular}
\end{center}
\end{table}
\clearpage

\section*{Supplementary Information}
\renewcommand{\thefigure}{S\arabic{figure}}
\renewcommand{\thetable}{S\arabic{table}}
\setcounter{figure}{0}
\setcounter{table}{0}
\setcounter{page}{1}

\begin{figure}
\includegraphics{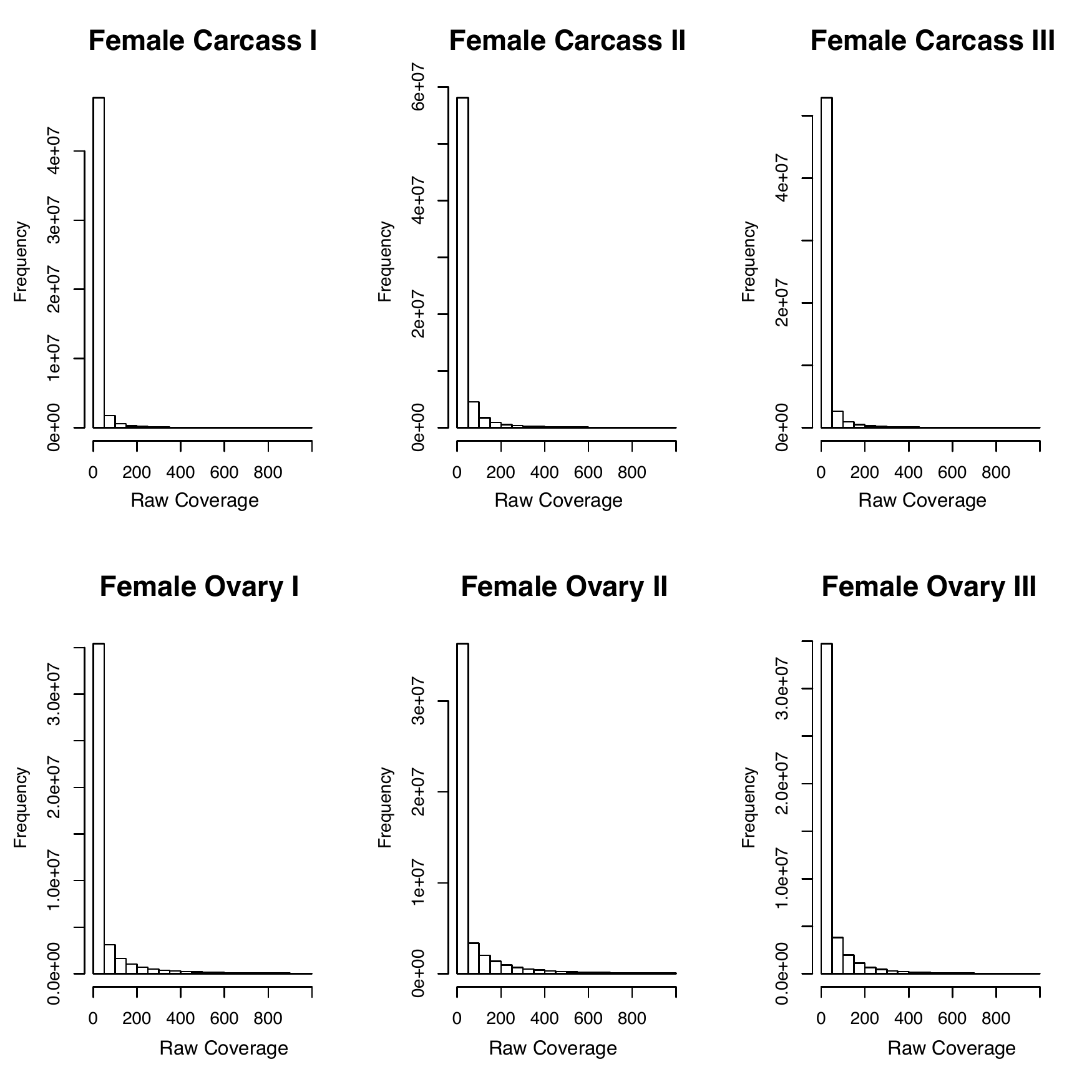}
\caption{\label{DsimFemCov} Raw coverage for sites with coverage sequencing depth between 1 and 1000 reads in RNA-seq data for replicates of female tissues in \Dsim.}

\end{figure}
\clearpage

\begin{figure}
\includegraphics{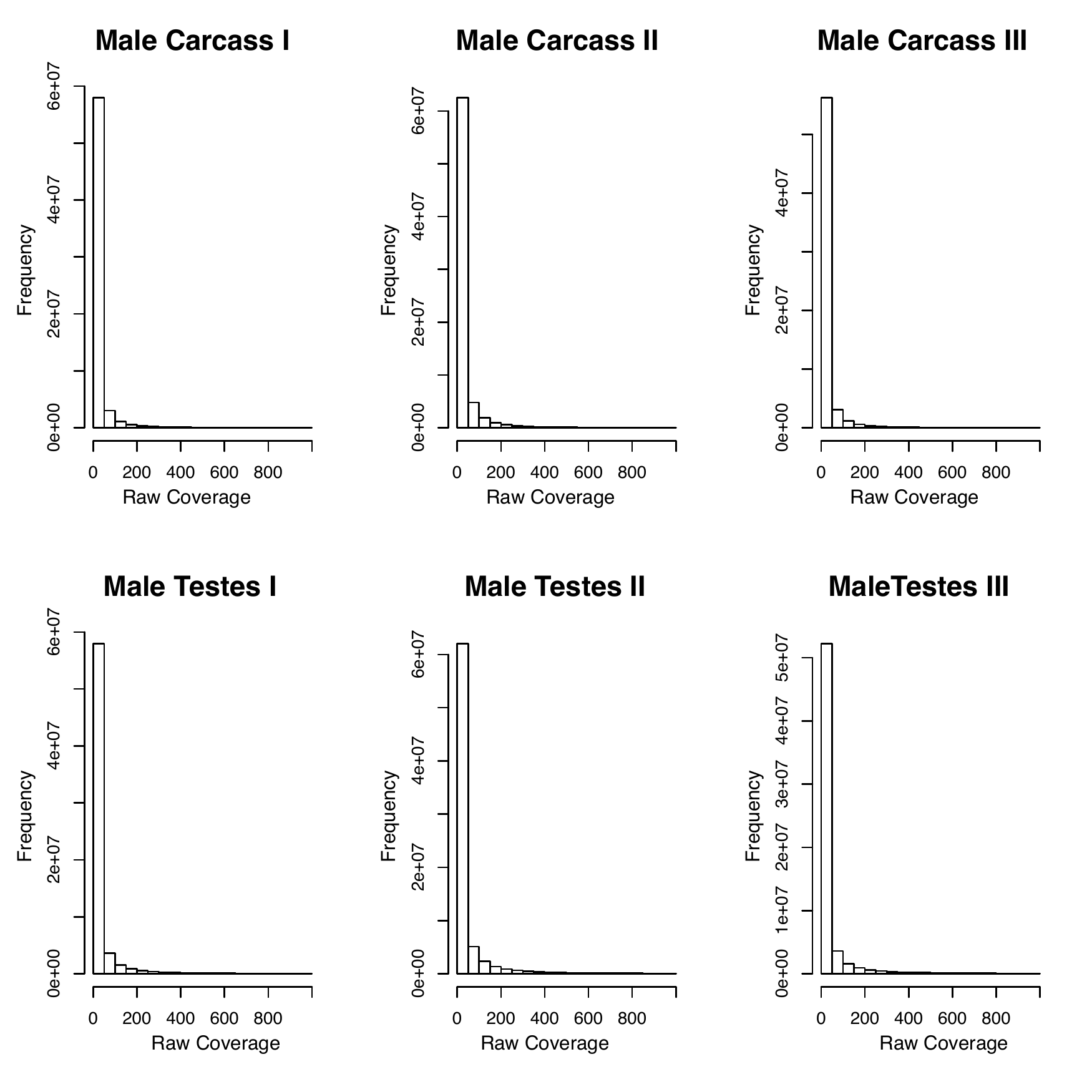}
\caption{\label{DsimMaleCov} Raw coverage for sites with coverage sequencing depth between 1 and 1000 reads in RNA-seq data for replicates of male tissues in \Dsim. }

\end{figure}
\clearpage

\begin{figure}
\includegraphics{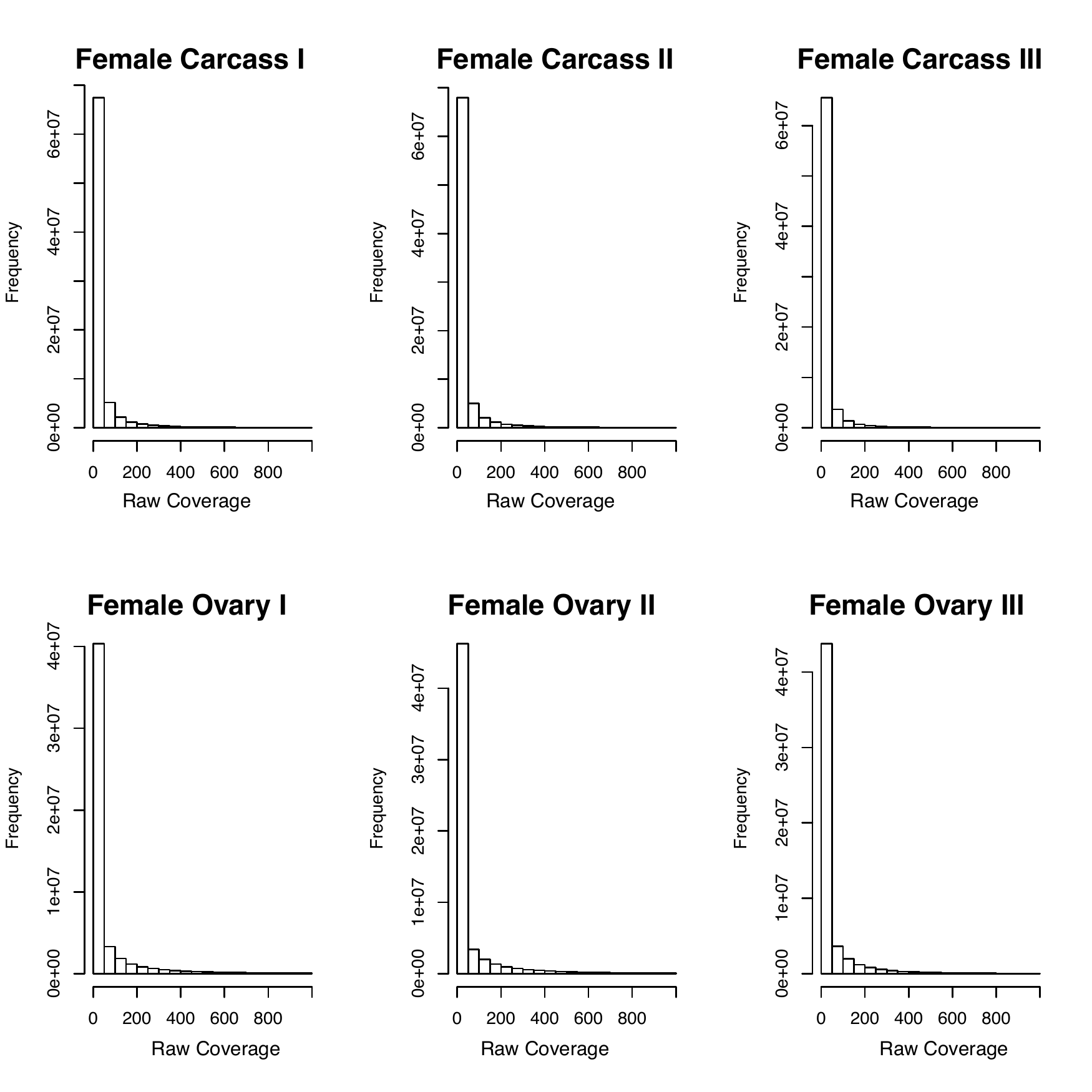}
\caption{\label{DyakFemCov} Raw coverage for sites with coverage sequencing depth between 1 and 1000 reads in RNA-seq data for replicates of female tissues in \Dyak.  }

\end{figure}
\clearpage

\begin{figure}
\includegraphics{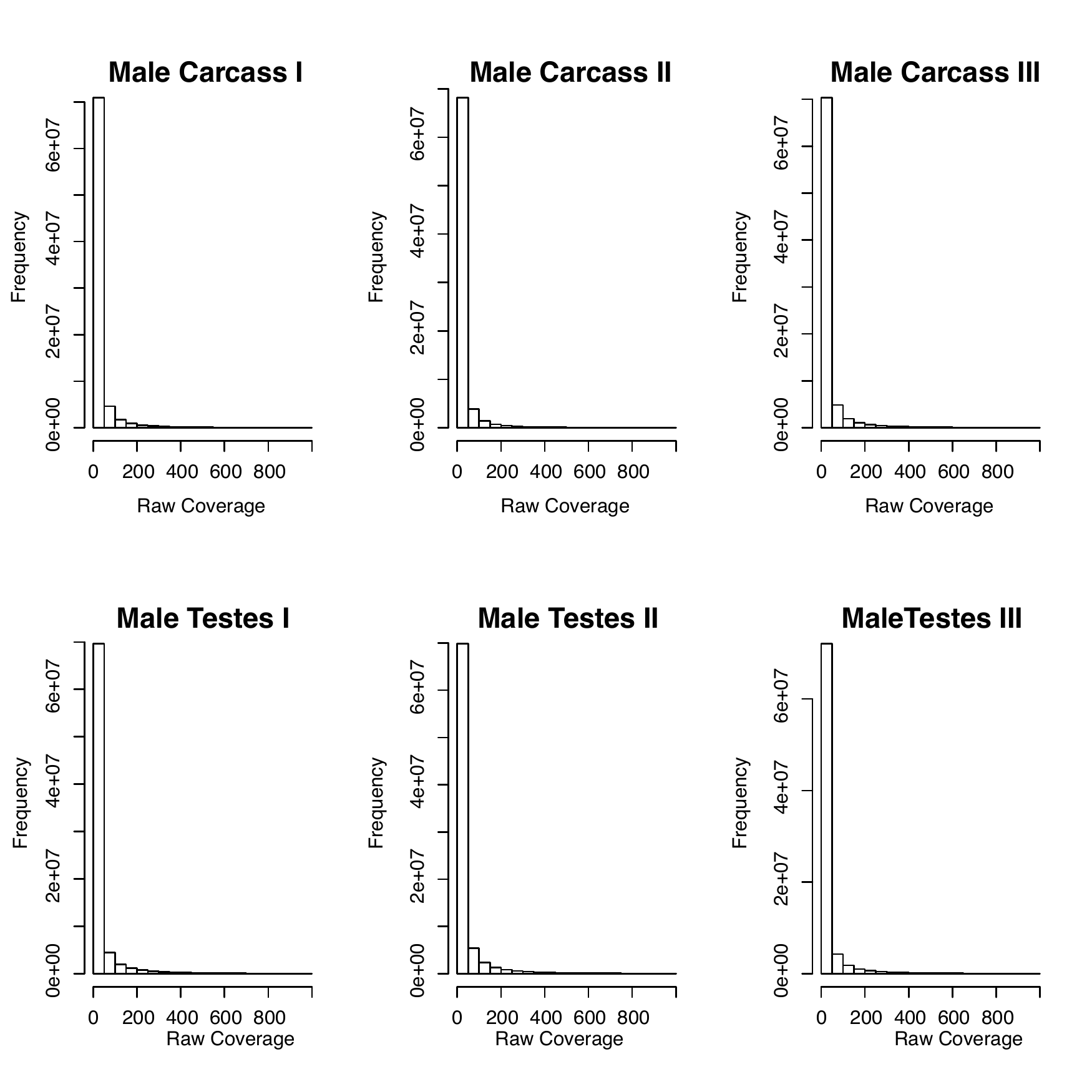}
\caption{\label{DyakMaleCov} Raw coverage for sites with coverage sequencing depth between 1 and 1000 reads in RNA-seq data for replicates of male tissues in \Dsim.  }

\end{figure}
\clearpage

\begin{figure}
\includegraphics{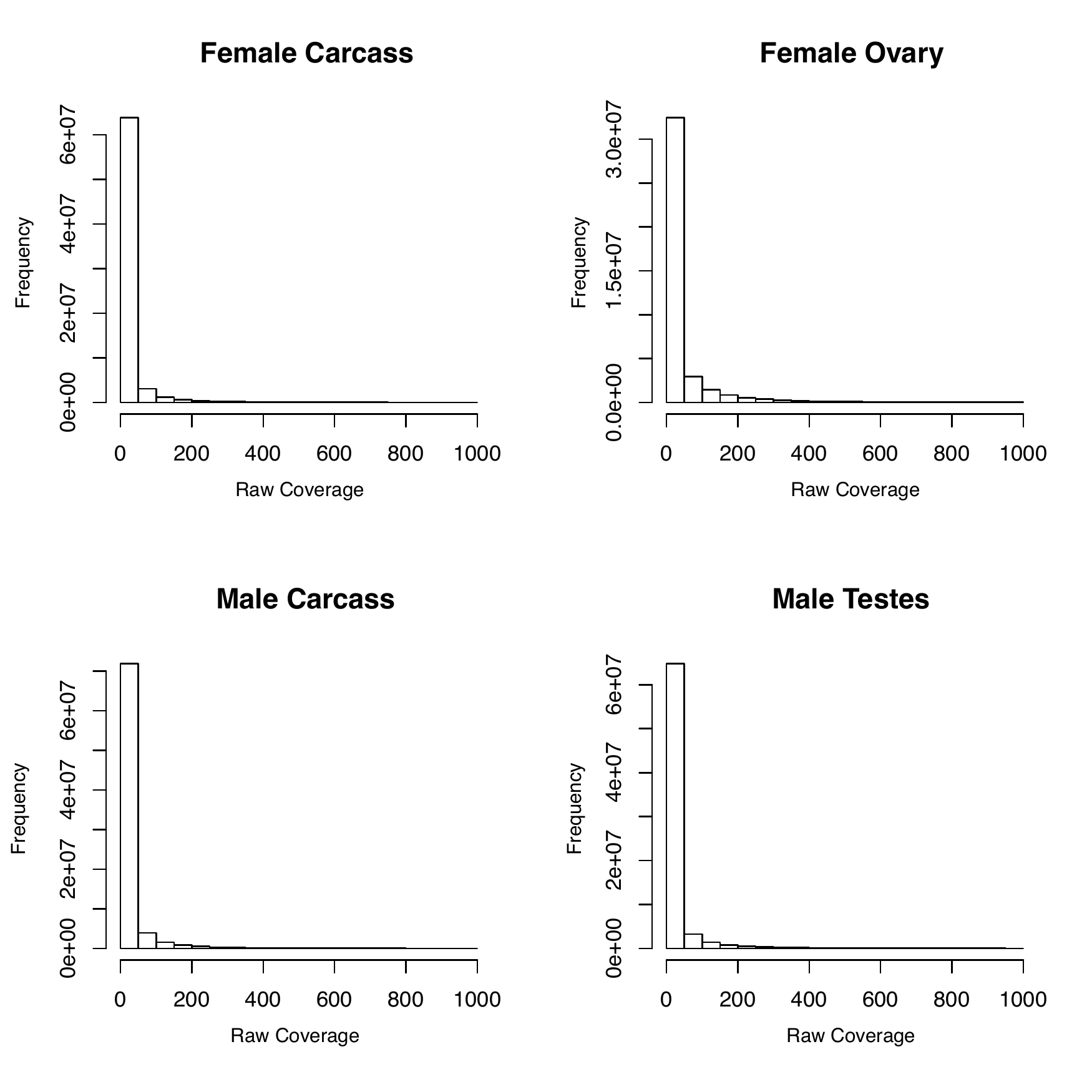}
\caption{\label{DanaCov} Raw coverage for sites with coverage sequencing depth between 1 and 1000 reads in RNA-seq data for tissues in \Dana.  }

\end{figure}
\clearpage

\begin{figure}
\includegraphics{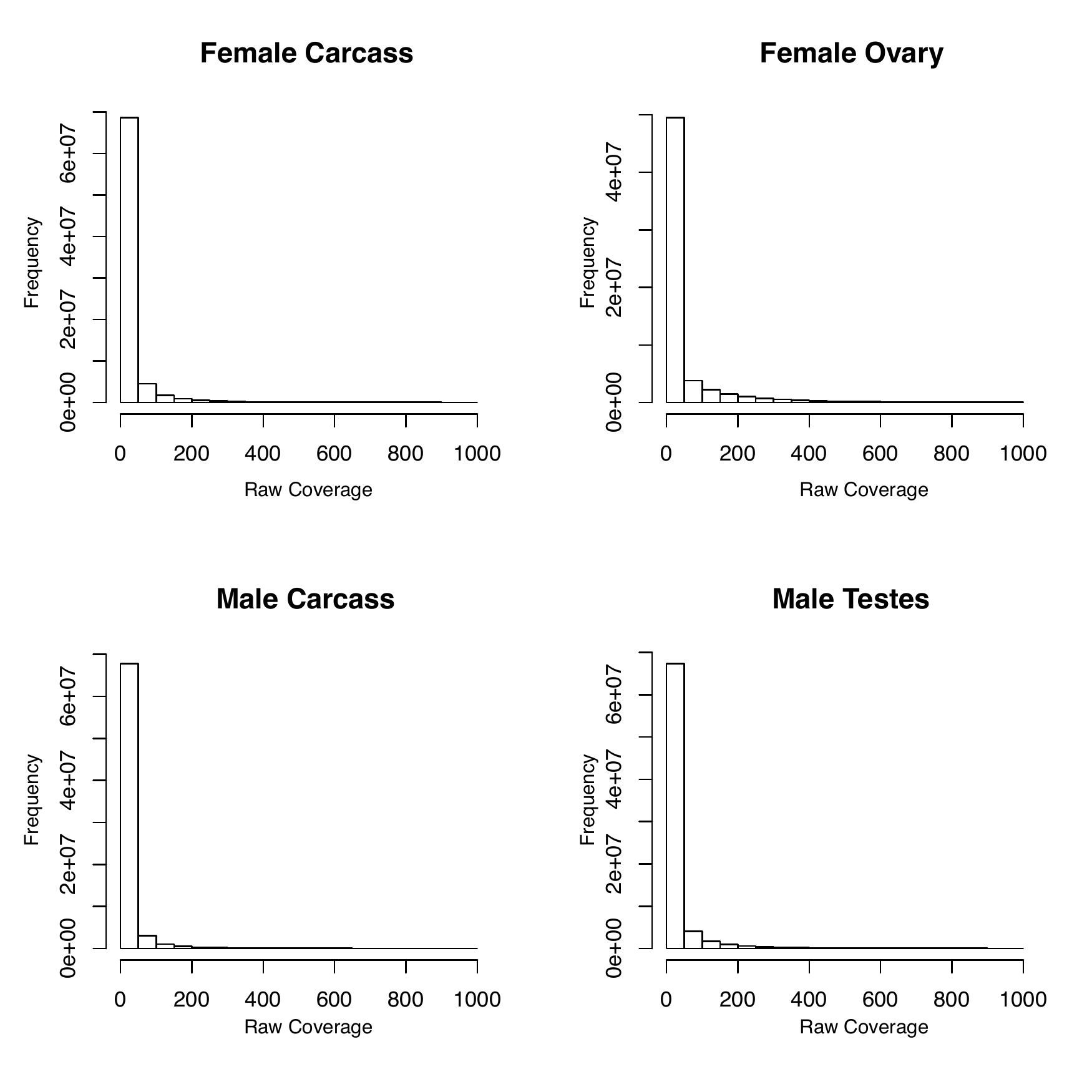}
\caption{\label{DmelCov} Raw coverage for sites with coverage sequencing depth between 1 and 1000 reads in RNA-seq data for tissues in \Dmel.  }

\end{figure}

\end{document}